\documentclass[conference,a4paper]{IEEEtran}
\usepackage[T1]{fontenc}
\usepackage[utf8]{inputenc}
\usepackage{cite}
\usepackage{amsmath,amssymb,amsfonts}
\usepackage{graphicx}
\usepackage{textcomp}
\usepackage{xcolor}
\usepackage{booktabs}
\usepackage{hyperref}
\hypersetup{
	hidelinks,
	plainpages   = false,
	pdftitle     = {{EfficientLEAF: A Faster LEarnable Audio Frontend of Questionable Use}},
	pdfsubject   = {{EUSIPCO-2022}},
	pdfauthor    = {{Jan Schlüter and Gerald Gutenbrunner}},
}
\begin{document}

\title{EfficientLEAF: A Faster LEarnable Audio Frontend of Questionable Use
}

\author{\IEEEauthorblockN{Jan Schlüter and Gerald Gutenbrunner}
\IEEEauthorblockA{\textit{Institute of Computational Perception} \\
\textit{Johannes Kepler University Linz, Austria}\\
\href{mailto:jan.schlueter@jku.at}{jan.schlueter@jku.at},
\href{mailto:gerald.gutenbrunner@gmail.com}{gerald.gutenbrunner@gmail.com}}
}

\maketitle

\begin{abstract}

In audio classification, differentiable auditory filterbanks with few parameters cover the middle ground between hard-coded spectrograms and raw audio.
LEAF \cite{leaf}, a Gabor-based filterbank combined with Per-Channel Energy Normalization (PCEN), has shown promising results, but is computationally expensive.
With inhomogeneous convolution kernel sizes and strides, and by replacing PCEN with better parallelizable operations, we can reach similar results more efficiently.
In experiments on six audio classification tasks, our frontend matches the accuracy of LEAF at 3\% of the cost, but both fail to consistently outperform a fixed mel filterbank.
The quest for learnable audio frontends is not solved.
\end{abstract}


\section{Introduction}
\label{sec:intro}

Deep-learning-based audio classification models typically operate on precomputed spectrograms -- this holds for Convolutional Neural Networks (CNNs) \cite{PANNs}, Recurrent Neural Networks \cite{graves2013}, and Transformers \cite{AST}. This places the burden of choosing optimal spectrogram settings for a task on the practitioner, who may decide not to tune these at all, possibly resulting in suboptimal performance.
Alternatively, models may be trained directly on raw audio samples. However, this gives the model much more free parameters, and only matches the performance of spectrogram-based models when given large quantities of training data \cite{sainath2015,pons2018}.
A solution in between these extremes is to apply a filterbank that is differentiable with respect to a small number of parameters, and learn these parameters along with the classifier.

A recent promising instance of the latter was proposed by Zeghidour et al.\ \cite{leaf} and called LEarnable Audio Frontend (LEAF). In their experiments, it outperforms earlier proposals by other authors when evaluated over a range of tasks in different audio domains (speech, music, environmental audio). However, as it is based on two convolutions (a Gabor filterbank and temporal pooling) with large windows and small strides, and normalization by a sequentially computed exponential moving average (Per-Channel Energy Normalization, PCEN \cite{pcen}), it is two orders of magnitude slower than typical spectrograms.

\begin{figure}
{
\small
\setlength{\tabcolsep}{1pt}
\begin{tabular}{llrl llrl llrl}
\multicolumn{4}{c}{filter 0}
& \multicolumn{4}{c}{filter 25}
& \multicolumn{4}{c}{filter 39} \\
\multicolumn{4}{r}{\includegraphics{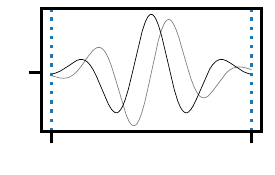}}
& \multicolumn{4}{r}{\includegraphics{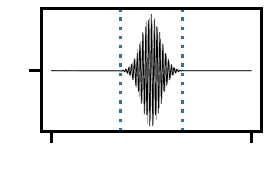}}
& \multicolumn{4}{r}{\includegraphics{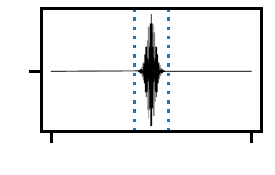}} \\
LEAF & conv & 401 // & 1   & \hphantom{LEAF} & conv & 401 // & 1   & \hphantom{LEAF} & conv & 401 // & 1\\
     & pool & 401 // & 160 && pool & 401 // & 160 && pool & 401 // & 160\\[.5em]
Ours & conv & 401 // & 40  && conv & 123 // & 4   && conv & 69 // & 1\\
     & pool &  11 // & 4   && pool & 101 // & 40  && pool & 401 // & 160\\
\end{tabular}
}
\caption{The original LEAF implementation convolves the input with filters of 401 samples at stride 1, followed by squared modulus and temporal pooling of 401 samples at stride 160.
We reduce filter lengths for higher bandwidths (dashed lines) and increase stride for lower center frequencies, adjusting pooling to approximate the original LEAF output.\label{fig:convsizes}}
\end{figure}

In this work, we propose two modifications of LEAF to improve computational performance by an order of magnitude without hindering trainability or hampering classification accuracy. Specifically, our modifications are:
\begin{itemize}
\item[--] We adapt convolution window sizes and strides dynamically for subsets of filters, giving nearly the same results with less computation (Figure~\ref{fig:convsizes}).
\item[--] We replace the sequentially computed normalization (PCEN) with a learnable logarithmic compression, temporal median subtraction and temporal batch normalization, all of which are parallelizable and thus faster to compute on Graphics Processing Units.
\end{itemize}
We evaluate our modifications against the original LEAF and fixed mel spectrograms on six tasks in three audio domains (speech, music, environmental audio).
Contrary to Zeghidour et al.\ \cite{leaf}, we find that \emph{none of the frontends has a clear advantage} over the others in terms of resulting accuracy.

The remainder of this paper is structured as follows: Section~\ref{sec:relwork} discusses related work, followed by an introduction of LEAF and our modifications in Section~\ref{sec:method}. In Section~\ref{sec:results}, we present experimental results. Section~\ref{sec:discussion} concludes the paper.

\section{Related Work}
\label{sec:relwork}

Existing attempts at implementing learnable filterbanks can be divided into categories based on two features of the filters: (1) The domain of operation, either time or frequency, and (2) the generation of coefficients, either given directly or produced by a parameterized function.
We will discuss selected examples of each category.

Sainath et al.\ \cite{sainath2013} learn the coefficients of frequency-domain filters initialized to a mel filterbank, constrained to their bandwidth at initialization, for speech recognition.
Cakir et al.\ \cite{cakir2016} remove this constraint, freely learning all coefficients for sound event detection.
In both cases, filters deviate from their initial triangular form, but stay close to a mel filterbank.

As examples of frequency-domain parametric filterbanks, Seki et al.\ \cite{seki2017} and Schlüter \cite[p.\,189]{schlueter2017} learn the center frequencies of Gaussian and triangular filters, respectively. Compared to directly learning coefficients, this reduces the number of learnable parameters and gives better interpretable filters, but is still based on a predefined, hand-tuned STFT.

Most freedom is attained by learning time-domain filter coefficients, as done e.g.\ by Palaz et al.\ \cite{palaz2013}, Tüske et al.\ \cite{tueske2014}, Sainath et al.\ \cite{sainath2015} or Zeghidour et al.\ \cite{zeghidour2018} for speech recognition.
Time-domain filters are often initialized to a mel \cite{zeghidour2018} or gammatone \cite{tueske2014,sainath2015} filterbank and followed by temporal pooling \cite{palaz2013,sainath2015,zeghidour2018}.
Early works failed to match performance of precomputed spectrograms \cite{palaz2013,tueske2014}, which only changed with larger datasets \cite{sainath2015} and models \cite{zeghidour2018}.

Parametric time-domain convolutions reduce learnable parameters and introduce inductive biases that may help training from limited datasets.
Existing work includes Sinc \cite{sincnet}, Sinc$^2$ \cite{loweimi2019}, Wavelet \cite{khan2018}, Gabor \cite{loweimi2019,doerfler2020}, Gammatone \cite{loweimi2019}, and Gammachirp \cite{espejo2020} filters with learnable center frequencies and/or bandwidths for spectral decomposition, and average pooling \cite{espejo2020}, max pooling \cite{sincnet,loweimi2019} and smoothing windows \cite{doerfler2020} for temporal downsampling.

All these works have in common that they are evaluated on a single dataset, often in the speech domain, leaving open whether a demonstrated advantage over fixed filterbanks transfers to other datasets, tasks or domains.
In contrast, LEAF \cite{leaf} was evaluated on eight tasks, and shown to outperform TF-banks \cite{zeghidour2018}, SincNet \cite{sincnet} and fixed mel filterbanks.

\section{Method}
\label{sec:method}

We will first describe LEAF \cite{leaf}, the starting point of our work, then detail our modifications of the filterbank and compression/normalization stages.

\subsection{LEAF}
\label{sec:leaf}

LEAF applies a Gabor filterbank, squared modulus, temporal averaging and subsampling, and a compression/normalization in sequence to an audio signal to obtain a time-frequency representation consumed by a classifier.
Stages are initialized to approximate a mel spectrogram, and optimized along with the classifier.
We will briefly describe each stage in the following.

\emph{Filterbank:} The first step is a convolution of the input signal with complex Gabor filters in the time domain. Given a center frequency $\nu$, inverse bandwidth $\sigma_c$ and odd filter size $C$, filter coefficients $c_t$ are computed as:
\[c_t = e^{i 2 \pi \nu t} \frac{1}{\sqrt{2 \pi} \sigma_c} e^{-\frac{t^2}{2 \sigma_c^2}} \quad \text{for } t \in \{-\frac{C-1}{2}, \dots, \frac{C-1}{2}\} \]
The coefficients are differentiable w.r.t.\ $\nu$ and $\sigma_c$. A filterbank of $N$ filters is thus parametrized by a vector of $N$ center frequencies and $N$ inverse bandwidths.

\emph{Squared modulus:} The $N$ convolved signals are squared elementwise, resulting in real-valued sequences.

\emph{Averaging:} Each sequence is convolved with a Gauss window. Given an inverse bandwidth $\sigma_p$ and odd pooling size $P$, window coefficients $p_t$ are computed as:
\[p_t = \frac{1}{\sqrt{2 \pi} \sigma_p} e^{-\frac{t^2}{2 \sigma_p^2}} \quad \text{for } t \in \{-\frac{P-1}{2}, \dots, \frac{P-1}{2}\} \]
Formally, the $N$ convolved sequences are then subsampled by keeping every $K$th value (practically, a strided convolution is applied that only computes every $K$th output). The averaging stage is parametrized by a vector of $N$ inverse bandwidths, such that pooling can be tuned separately for each filter.

\emph{Compression/Normalization:} Finally, Per-Channel Energy Normalization (PCEN \cite{pcen}) is applied to each sequence. Given $\epsilon$, $\alpha$, $\delta$, $r$ and $s$, and denoting the input sequence as $x_t$, the output sequence $y_t$ is:
\[ y_t = \left( \frac{x_t}{\left(\epsilon + m_t\right)^{\alpha}} + \delta \right)^r - \delta^r, \]
where $m_t$ is computed using a simple infinite impulse response (IIR) filter:
\[ m_0 = x_0, \quad m_t = (1 - s) m_{t-1} + s\,x_t \]
This process is applied separately to each of the $N$ sequences, using a separate set of learnable parameters for each (except for $\epsilon$, which is fixed).
The result is a division of each frequency band by its long-term past magnitude (sequence $m_t$), and a nonlinear compression by raising to the power of $r$.
Wang et al.\ \cite{pcen} learned the logarithm of $\alpha$, $\delta$, $r$; Zeghidour et al.\ \cite{leaf} instead learn the inverse of $r$ and enforce $\alpha \leq 1$, $r \leq 1$.

\subsection{EfficientLEAF}
\label{sec:efficientleaf}

We are now ready to discuss our changes to the LEAF filterbank, pooling and normalization/compression stages.

\emph{Filterbank:} LEAF initializes filters to a mel scale, with roughly logarithmically increasing center frequencies and bandwidths.
With increasing bandwidth, filter energy concentrates in fewer coefficients (see Figure~\ref{fig:convsizes}). We can thus save computations by truncating the filter. Specifically, we compute a filter size $\hat C = b \, \sigma_c$ and round up to the next odd integer, where $\sigma_c$ is the inverse bandwidth and $b$ can be tuned to trade accuracy for computation. Khan et al.\ \cite{khan2018} proposed to do so for a complete filterbank, here we compute $C$ for each filter separately.
With decreasing center frequency, filter responses are smoother over time, and change less from sample to sample. We can thus save computations by increasing the convolution stride. Specifically, we compute $\hat L = d\,\pi / \nu$ and round down to the next divisor of the pooling stride $P$, where $\nu$ is the center frequency and $d$ can be tuned to trade accuracy for computation.
Since convolution implementations profit from applying multiple same-sized filters at once, we group adjacent filters and pick the largest filter size and smallest stride per group. The number of groups $g$, ideally a divisor of $N$, becomes another hyperparameter.

\emph{Pooling:} Both the pooling stride $P$ and the pooling scale $\sigma_p$ need to be divided by the convolution stride $L$ to match results of the original LEAF.
%
Figure~\ref{fig:convsizes} gives the resulting window sizes and strides for matching the default settings of LEAF, with $b=4.75$ chosen to reproduce a maximal window size of $401$ at initialization, $g=4$, and $d=1$ chosen conservatively.

\emph{Normalization/Compression:} The sequential computation of the exponential moving average in PCEN is not suited well for massively parallel hardware. We replicate some of its effects by different means.
As a first step, we compute $y_t = \log(1 + 10^a\,x_t)$, where $a$ is a separate learnable parameter for each frequency band. This results in a nonlinear compression similar to exponentiation by $r$.
PCEN's division by an exponential moving average levels out different impulse responses of recording devices, or static noise floors. As we applied an (approximate) logarithm, we require subtraction instead of division. To avoid the exponential moving average, we subtract the median over the sequence instead (separately per frequency band). As this improves performance for some tasks only, reducing it for others, we keep the original sequence as a second input channel.
Finally, we normalize the sequence with batch normalization over time, using separate learnable parameters per frequency band and channel.

\section{Experiments and Results}
\label{sec:results}

We can now empirically compare EfficientLEAF to LEAF, and to a fixed mel filterbank. We will first introduce the datasets used, then describe training and model settings, and finally present results for three experiments: Our main comparison, a hyperparameter optimization of EfficientLEAF, and an extension to longer input sequences.

\subsection{Datasets}
\label{sec:datasets}

For our experiments, we employ five datasets with six tasks:\\
-- \emph{SpeechCommands}: one-second recordings of 35 spoken commands; 84843 training, 9981 validation, 11005 test\\
-- \emph{VoxForge}: variable-length recordings in 6 languages; 128594 training, 44119 validation, 30136 test\\
-- \emph{Crema-D}: variable-length recordings displaying 6 emotions; 5144 training, 738 validation, 1555 test\\
-- \emph{NSynth}: 4-second recordings of 11 instruments in 128 pitches; 289205 training, 12678 validation, 4096 test\\
-- \emph{BirdCLEF 2021}: variable-length recordings of 397 bird species; 40836 training, 5637 validation, 16401 test\\
If no split was published along with the data, we use the one from tensorflow\_datasets\footnote{\url{https://tensorflow.org/datasets}, accessed June 12, 2022}. Unfortunately, Zeghidour et al.\ \cite{leaf} did not publish their splits, and we could not reproduce any.

\subsection{Settings}
\label{sec:settings}

We set up LEAF to match \cite{leaf}: An input sample rate of 16\,kHz, 40 filters initialized with a mel scale from 60\,Hz (lower bound of first filter) to 7800\,Hz (upper bound of last filter), a convolution and pooling window size of 401 samples, and a pooling stride of 160 samples. Pooling scales $\sigma_p$ are initialized to $0.4$.
PCEN is initialized with $\alpha=0.96$, $s=0.04$, $\delta=2$, $r=0.5$ and has $\epsilon=10^{-12}$.
For EfficientLEAF, we set $b=4.75$, $d=1$, $g=4$, $a=5$ as a close match to LEAF, but we perform a parameter search in our second experiment.

For classification, we follow \cite{leaf} and add an EfficientNet-B0 \cite{efficientnet} backbone with global max pooling instead of global average pooling, and a single linear classification layer.

To train the model, Zeghidour et al.\ \cite{leaf} used ADAM with mini-batches of 256 randomly chosen one-second excerpts, and ran 1 million updates at a constant learning rate of $10^{-4}$.
This amounts to thousands of epochs depending on the dataset, and a constant learning rate seems suboptimal.
Instead, we start with an initial learning rate of $10^{-3}$, reduce it by a factor of ten when the validation loss does not improve for ten consecutive epochs, and stop training when the learning rate falls below $10^{-5}$.
This improves results for all frontends.

At test time, we compute predictions for non-overlapping one-second excerpts and average logits per recording, following \cite{leaf} except that final incomplete excerpts are dropped, not padded (which skews results as no padding occurs in training).

\subsection{Model Comparison}
\label{sec:comparison}

\begin{table*}[t!]
\begin{minipage}{.65\linewidth}
\caption{Throughput of audio frontend in examples per second (on one-second excerpts) and accuracy on six tasks (mean $\pm$ std.\ dev.\ over three runs), for five combinations of filterbank and compression/normalization. (*: parameters fixed, not learned)\label{tab:comparison}}
{
\begin{tabular}{llllll}\toprule
Filterbank & Gabor & Gabor 4G & Gabor 4G & Gabor 8G-opt & STFT-Mel*\\
Compression & PCEN & PCEN & L-M-TBN & L-M-TBN & L-M-TBN*\\
\midrule
Throughput & 250 & 742 & 776 & 9251 & 85367\\
\midrule
SpeechCommands & 95.1 $\pm$ 0.3 & 95.1 $\pm$ 0.1 & \textbf{95.3} $\pm$ 0.2 & 95.2 $\pm$ 0.1 & 95.1 $\pm$ 0.2\\
VoxForge       & \textbf{91.5} $\pm$ 0.4 & 91.4 $\pm$ 0.9 & 86.5 $\pm$ 0.9 & 86.6 $\pm$ 1.0 & 85.6 $\pm$ 0.6\\
Crema-D        & 50.2 $\pm$ 2.3 & 50.0 $\pm$ 2.6 & 58.0 $\pm$ 2.8 & \textbf{60.2} $\pm$ 0.8 & 58.8 $\pm$ 3.2\\
NSynth Instr.  & 69.2 $\pm$ 0.2 & 68.3 $\pm$ 1.2 & 70.4 $\pm$ 0.5 & 71.7 $\pm$ 0.6 & \textbf{72.1} $\pm$ 0.7\\
NSynth Pitch   & 92.2 $\pm$ 0.1 & 92.1 $\pm$ 0.1 & \textbf{92.7} $\pm$ 0.2 & 92.4 $\pm$ 0.1 & 91.9 $\pm$ 0.3\\
BirdCLEF 2021  & 42.3 $\pm$ 0.7 & 42.3 $\pm$ 0.8 & 42.0 $\pm$ 0.1 & \textbf{42.9} $\pm$ 0.1 & 39.9 $\pm$ 1.9\\
\bottomrule
\end{tabular}
}
\end{minipage}
\hfill
\begin{minipage}{.29\linewidth}
\caption{Throughput and accuracy for the first, third and fourth model from Table~\ref{tab:comparison} on BirdCLEF 2021, trained on 8- or 16-second excerpts.\label{tab:longer}}
\vspace{-1.35mm}
{
\begin{tabular}{lrr}\toprule
length (s) & 8 & 16\\
batchsize & 32 & 16\\
\midrule
\#1 thrpt. & 27 & 12\\
\#1 acc. & 71.9 $\pm$ 0.4 & 69.6 $\pm$ 0.4\\
\midrule
\#3 thrpt. & 95 & 48\\
\#3 acc. & 71.4 $\pm$ 0.9 & 66.0 $\pm$ 2.4\\
\midrule
\#4 thrpt. & 1053 & 516\\
\#4 acc. & 72.2 $\pm$ 0.3 & 69.4 $\pm$ 0.3\\
\bottomrule
\end{tabular}
}
\end{minipage}
\end{table*}

In our first set of experiments, we compare a set of models on the six tasks.
Starting with the original LEAF, we first replace the filterbank and pooling with our grouped version, then replace PCEN with our combination of log compression, median filtering and temporal batch normalization (``L-M-TBN''). Finally, we replace the filterbank and pooling with a fixed STFT-based mel filterbank (also using a window size of 401 and stride 160) and hold log compression fixed.

Table~\ref{tab:comparison} lists the results (ignore the second to last column for now).
Focusing on throughput (forward + backprop), we see that the grouped Gabor filterbank at its conservative settings is 3x as fast, and replacing PCEN gives another 5\% (this will be more pronounced for longer input sequences). Fixed mel spectrograms are 100x faster and could even be precomputed.
In terms of accuracy, there seems to be a consistent decline when replacing PCEN for VoxForge. However, results for VoxForge are either extremely sensitive to the split, or models are overfitting: On the validation set, accuracies behave inversely, improving from 74.2\% for LEAF to 79.8\% for a fixed mel filterbank. The poor performance of PCEN-based frontends on Crema-D, the smallest dataset, warrants investigation.

\subsection{Hyperparameter Optimization}
\label{sec:hyperparams}

EfficientLEAF has three parameters affecting its efficiency and accuracy: The convolution window size factor $b$, convolution stride factor $d$, and number of groups $g$. We perform a grid search with $b \in \{2, 4.75, 6\}$, $d \in \{1, 2, 3, 8, 16\}$ and $g \in \{2, 4, 8, 10\}$, doing 3 training runs on SpeechCommands each.
For space constraints, we can only summarize results.
For almost all settings, $g=8$ is the fastest.
$b=2$ slightly deteriorates results, $b=6$ is only marginally slower than $b=4.75$.
$d$ scales computational speed almost linearly, without affecting results on this task in the range of considered values. This is in line with Dörfler et al.\ \cite{doerfler2020}, who use a stride of $21$ for a sample rate of 22050\,Hz.
In Table~\ref{tab:comparison}, the previous to last column shows results with $g=8$, $b=6$ and $d=16$, which match the more conservative settings of $g=4$, $b=4.75$ and $d=1$ at much better efficiency.

\subsection{Longer Input Sequences}
\label{sec:longer}

Following \cite{leaf}, all results discussed so far were obtained by training and evaluating on one-second audio excerpts. This recipe is not applicable to every audio classification task. For example, for weakly-labeled recordings, not every excerpt will be discriminative, as is the case for the BirdCLEF 2021 data. In this setting, it will be necessary to train on longer excerpts.

Table~\ref{tab:longer} shows results for training and evaluating the original LEAF, the default EfficientLEAF and optimized EfficientLEAF on either 8-second or 16-second excerpts, for BirdCLEF 2021.
Two observations are important: (1) longer excerpts indeed perform dramatically better, and (2) while EfficientLEAF throughput scales inversely linearly with input length, LEAF is put at a larger disadvantage, increasing the gap in throughput. This is due to PCEN: As it has to process each item sequentially, a batch of 32 8-second excerpts allows fewer parallel computations than a batch of 256 1-second excerpts, stalling the GTX 1080 Ti used for testing.

\section{Discussion}
\label{sec:discussion}

We have demonstrated that LEAF \cite{leaf} can be modified to improve computational efficiency, especially for long input sequences, without impacting accuracy on downstream tasks.
We also found that LEAF may not be needed:
Our deviation from Zeghidour et~al.~\cite{leaf} in training and inference (Sec.~\ref{sec:settings}) and compression (Sec.~\ref{sec:efficientleaf}) improved results, but also narrowed the advantage of LEAF over a fixed mel filterbank.
Whether and why LEAF is beneficial will require further scrutinization and experiments, and maybe our implementation (\href{https://github.com/CPJKU/EfficientLEAF}{github.com/CPJKU/EfficientLEAF}) can speed up this process.

Regarding EfficientLEAF, an interesting feature has not been explored yet: Since convolution window sizes are chosen dynamically, it could learn to analyze lower frequencies than would be permitted by a predefined window size, or be initialized to cover a much wider range of frequencies than affordable with a fixed window.

Finally, during experimentation, we observed that learned center frequencies and bandwidths do not deviate much from their initial values (in line with \cite[A.3]{leaf}). As in \cite[189--190]{schlueter2017}, we tried increasing the frontend learning rate. This indeed allows some frontend parameters to converge, but reduces classification performance, asking for a better solution.


\bibliographystyle{IEEEtran}
\bibliography{references}

\end{document}